\documentclass[11pt]{article}

\usepackage[margin=1in]{geometry}
\usepackage{siunitx}
\usepackage{upgreek}
\usepackage{graphicx}
\usepackage{amsmath}
\usepackage{amssymb}
\usepackage{xfrac}
\usepackage{xcolor}
\usepackage{soul}
\usepackage{pdfpages}
\usepackage{bm}

\usepackage{mathptmx}

\usepackage[super,comma,sort&compress]{natbib}
\usepackage{titlesec}
\titleformat{\section}{\Large\bfseries}{\thesection.}{1em}{}[]
\titleformat{\subsection}{\normalsize\bfseries}{\thesubsection.}{1em}{}[]
\titleformat{\subsubsection}{\normalsize\itshape}{\thesubsubsection.}{1em}{}[]
\titlespacing{\section}{0pt}{5pt}{0pt}[0pt]
\titlespacing{\subsection}{0pt}{3pt}{0pt}[0pt] % Indent set to 10 pt below
\titlespacing{\subsubsection}{0pt}{0pt}{0pt}[0pt] % Indent set to 10 pt below

\usepackage[font=footnotesize,labelfont=bf]{caption} 

\usepackage{float}
\usepackage[color=blue!15, size=footnotesize]{todonotes}
\usepackage{xcolor}
\begin{document}

\begin{center}

\Large
\textbf{Traction and Stress Control Formation and Motion of +1/2 Topological Defects in Epithelial Cell Monolayers}

\normalsize

Pradip K. Bera$^{1,*}$, Molly McCord$^{1,2,*}$, Jun Zhang$^{2,3}$, Jacob Notbohm$^{1,2,3,\dagger}$

$^1$ Department of Mechanical Engineering, University of Wisconsin--Madison, Madison, Wisconsin, 53706, USA \\
$^2$ Biophysics Program, University of Wisconsin--Madison, Madison, Wisconsin, 53706, USA \\
$^3$ Department of Nuclear Engineering and Engineering Physics, University of Wisconsin--Madison, Madison, Wisconsin, 53706, USA \\
$^*$ These authors contributed equally. \\
$^\dagger$ Correspondence: jknotbohm@wisc.edu

\end{center}

\vspace{11pt}

\section*{Abstract}

In confluent cell monolayers, patterns of cell forces and motion are systematically altered near topological defects in cell shape. In turn, defects have been proposed to alter cell density, extrusion, and invasion, but it remains unclear how the defects form and how they affect cell forces and motion. Here, we studied +1/2 defects, and, in contrast to prior studies, we observed the concurrent occurrence of both tail-to-head and head-to-tail defect motion in the same cell monolayer. We quantified the cell velocities, the tractions at the cell-substrate interface, and the stresses within the cell layer near +1/2 defects. Results revealed that both traction and stress are sources of activity and dissipation within the epithelial cell monolayer, with the direction of motion of +1/2 defects depending on whether energy is injected by stresses or tractions. Interestingly, patterns of motion, traction, stress, and energy injection near +1/2 defects existed before defect formation, suggesting that defects form as a result of spatially coordinated patterns in cell forces and motion. These findings introduce a new focus, on coordinated patterns of force and motion that lead to defect formation and motion.

\section*{Key Words}

collective cell migration, topological defects, traction force microscopy, energy injection and dissipation, active matter

\section*{Introduction}

In embryonic morphogenesis, collective cell migration, and cell extrusion, cells often align the long axes of their bodies with those of their neighbors, producing nematic order. \cite{reffay2011orientation, saw2017topological, maroudas2021topological, guillamat2022integer} Flaws in the order, called topological defects, can affect the cell layer in important ways, such as by altering the local cell density, inducing formation of holes, causing extrusion of cells from the cell layer, and altering rates of cancer cell invasion.\cite{kawaguchi2017topological, saw2017topological, copenhagen2021topological, zhang2021topological} Consistent with these phenomena, patterns of cell velocities are systematically altered near the defects.\cite{duclos2017topological, kawaguchi2017topological, saw2017topological, blanch2018turbulent, guillamat2022integer, balasubramaniam2021investigating} Additionally the stress state within the monolayer is altered near topological defects, with the presence of clear local gradients in intercellular stresses.\cite{saw2017topological, balasubramaniam2021investigating, letoquin2024mechanical} 

Given the systematic patterns of cell velocity, density, and stresses occurring at defects, topological defects provide a useful means of studying the relationships between cell shape, velocity, and force production. However, it is not yet clear how the defects in nematic order form and how exactly they are related to the cell forces and the collective migration. It is commonly assumed that cell monolayers are well described by the theory of active nematic materials, in which the amount of nematic order results from a competition between the tendency for cells to align with their neighbors and the magnitude of active stresses produced by the cells.\cite{marenduzzo2007steady, marchetti2013hydrodynamics, giomi2014defect} In this description, defects form as the active stresses increase relative to the tendency for alignment between neighboring cells. Theories often assume that the force generating machinery within each cell aligns with the major axis of the cell body.\cite{kawaguchi2017topological, saw2017topological, copenhagen2021topological, balasubramaniam2021investigating}  Following this assumption, one can predict cell velocity patterns near defects, but, paradoxically, epithelial cells that produce contractile (tensile) stresses have been observed to move opposite to predictions, as if their active stresses were extensile (pushing).\cite{saw2017topological, blanch2018turbulent, balasubramaniam2021investigating} The cause of this discrepancy has not yet been determined, but it may be that prior studies did not fully consider the forces at the cell-cell and cell-substrate interfaces.

When applied to cell monolayers, theories based on active nematic materials have typically assumed that the traction between the cell and the substrate acts as a damping friction that impedes motion,\cite{thampi2014active, duclos2018spontaneous, kumar2018tunable, shankar2018defect, angheluta2021role} with some theory being supported by experimental data.\cite{kawaguchi2017topological, copenhagen2021topological, han2025local} For epithelial cells, it is not obvious that cell-substrate traction would act as a passive friction, especially given prior observations that tractions can have strong effects on cell motion and aspect ratio. \cite{saraswathibhatla2020tractions} Some models for active nematics have considered that traction may have an additional part that propels the cell forward,\cite{vafa2021fluctuations, killeen2022polar} as have other models for collective cell migration that do not consider the cell orientation field.\cite{lee2011crawling, basan2013alignment, notbohm2016cellular, bi2016motility} Additionally, some experiments in epithelial cell monolayers have observed situations in which tractions align with, rather than against, the cell velocity.\cite{kim2013propulsion, notbohm2016cellular, saraswathibhatla2021} As alignment of traction and velocity would inject energy into the system, the observations of alignment between traction and velocity indicates that cell-substrate tractions are another potential source of activity. This complicates understanding of what parameters control nematic order and formation of defects, as now the tendency for alignment between neighboring cells would be balanced against both the active stresses within the cell layer and the active tractions at the cell-substrate interface. Hence, further investigation is required to uncover the relationships between nematic defects, the stresses, the tractions, and the collective cell motion.

Here, we study the forces leading to formation and motion of +1/2 topological defects in islands of Madin Darby canine kidney (MDCK) epithelial cells. We quantify the stresses within the cell layer and the tractions at the cell-substrate interface. We observe that some defects move in the tail-to-head direction, consistent with prior observations for this cell type.  \cite{balasubramaniam2021investigating} Interestingly, other defects in the same monolayer move in the opposite direction, from the head to the tail. Through detailed analysis of the forces and kinematics in these experiments, we find that the tail-to-head defects move against gradients in intercellular stress and with the direction of traction. This observation identifies that traction can act as a driving force to propel +1/2 defects against gradients in stress. To combine force and displacement into a single parameter, we quantify the power densities associated with stresses and traction, showing that for tail-to-head moving defects, the energy for cell motion comes primarily through the cell-substrate tractions, whereas for head-to-tail defects, the energy comes from transmission of stresses applied by neighboring cells. Finally, we show that patterns of motion, stresses, tractions, and energy injection associated with +1/2 defects existed in the cell layer before defect formation, suggesting that formation and motion of nematic defects may result from patterns in force and motion occurring before the defects form.

\section*{Results}

\subsection*{Motion of Cells near Topological Defects}

We seeded MDCK cells in islands of 1 mm diameter on polyacrylamide substrates of Young's modulus 6 kPa, as in our prior work.\cite{notbohm2016cellular, saraswathibhatla2020tractions} Cells were seeded to confluence and imaged at a relatively low density of approximately 2400 cells/mm$^2$, which results in a larger aspect ratio compared to higher densities. \cite{saraswathibhatla2020tractions} In some experiments, we modulated the cell forces using CN03 (2 {\textmu}g/mL), which activates Rho thereby increasing actomyosin contraction, and, separately, using CN02 (40 ng/ml), which activates Rac/Cdc42 activity thereby perturbing organization of actin within the cytoskeleton. Both treatments increased the average traction and the cell contractile stresses but did not alter the average cell migration speed (Supplemental Fig. S1). This finding is consistent with prior studies that showed no effect of CN03 on average cell speed.\cite{saraswathibhatla2020tractions, jipp2024cell} Following prior efforts, \cite{saw2017topological, kawaguchi2017topological, copenhagen2021topological} we imaged the cell layers by optical microscopy over time and identified topological defects. From an image of the cell layer (Fig. 1a), the orientation field was measured, and topological defects were identified by singularity points. \cite{kawaguchi2017topological, balasubramaniam2021investigating, saw2017topological, zhang2021topological, zhang2022} The topological defect charge, defined as the rotation experienced by the orientation field along a path encircling the defect, was either +1/2 or -1/2, represented by comet and tripod symbols, respectively (Fig. 1b). 

\begin{figure}[th!]
\includegraphics[width=6in]{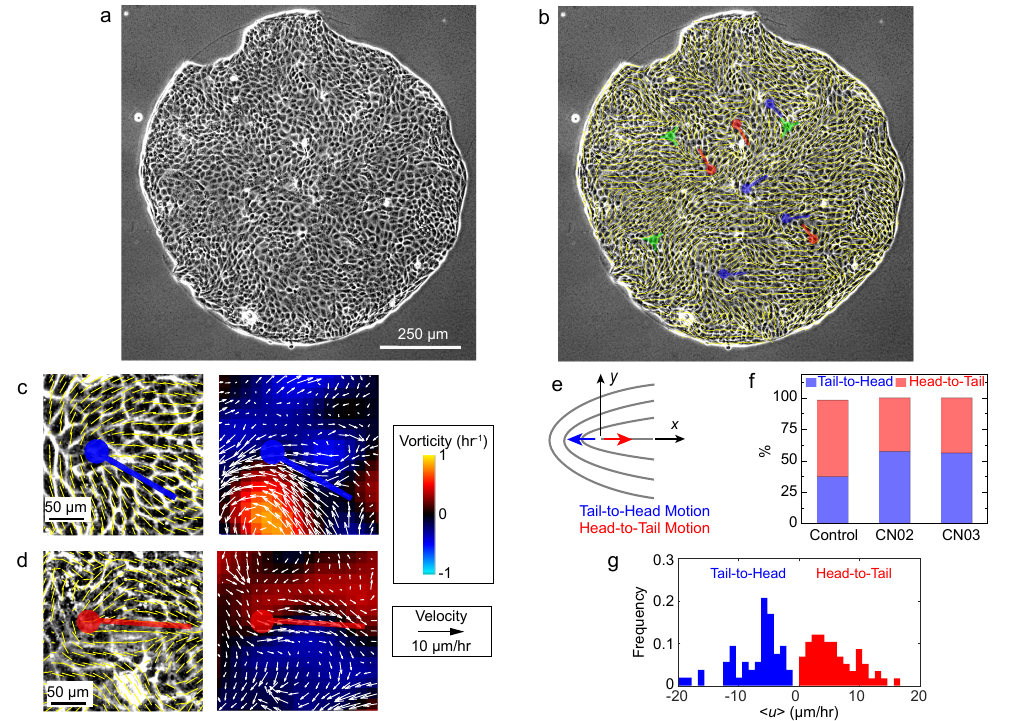}
\caption{
Topological defects in MDCK cell layers. 
(a) Phase contrast image of a confluent MDCK cell island. 
(b) Cell orientation plot with yellow lines indicating local orientation. +1/2 and -1/2 defects are indicated by the comet tails and tripods, respectively. Scale bar is 250 {\textmu}m.
(c, d) Representative +1/2 defects in MDCK cell layers moving in the tail-to-head direction (c) and head-to-tail direction (d). 
Yellow lines indicate cell orientation, white arrows indicate cell velocity, and color maps indicate vorticity. 
Note that the vorticity has opposite signs on either sides of the defect tails.
(e) Local coordinate system used to analyze all +1/2 defects having the origin $(x=0,y=0)$ located at the defect head and the positive $x$ direction pointing along the defect tail. Arrows show motion of the defect in the  tail-to-head (blue) and head-to-tail (red) directions.
(f) Fraction of defects with tail-to-head and head-to-tail motion in cell layers under the different treatments.
(g) Distribution of average $x$ component of cell velocity ($<\!v_x\!>$) for 111 identified +1/2 defects. 
For each defect, $v_x$ is averaged over cells located in the tail region ($0 < x < 100$ {\textmu}m, $y=0$). 
Out of 111 +1/2 defects, there were 58 head-to-tail and 53 tail-to-head moving defects.
}
\end{figure}

As isolated -1/2 defects are not expected to move due to their three-fold symmetry, we focused on +1/2 defects for this study. Identified defects were verified by quantifying the corresponding cell velocity and vorticity fields using image correlation \cite{bar2015fast} and comparing to theoretical predictions, namely the +1/2 defects are predicted to have a pair of vortices on either side of the defect's tail. For +1/2 defects, the average cell velocity can be either in the tail-to-head or head-to-tail direction, with the direction thought to depend on whether the system is, respectively, extensile, wherein the cells create pressure and push along their long axis, or contractile, wherein the cells produce tension and pull along their long axis. Near the +1/2 defects, the cells followed the predicted patterns of vorticity and velocity, and, interestingly, both tail-to-head and head-to-tail cell velocities were observed in all cell islands (Fig. 1c,d, Supplemental Fig. S2). After quantifying cell velocities, we tracked the motion of the +1/2 defects themselves. The direction of motion of the defect (tail-to-head or head-to-tail) matched that of the cell velocity (Supplemental Fig. S2 and Videos 1-3). Prior studies \cite{saw2017topological, blanch2018turbulent, balasubramaniam2021investigating, zhang2022} interpreted the direction of defect motion as indicating extensile or contractile behavior. The use of defect velocity to distinguish between extensile and contractile has been recently drawn into question, \cite{nejad2024stress} however, so hereafter we use the language ``tail-to-head'' and ``head-to-tail'' to describe cell migration and defect motion.

Prior reports with epithelial cell types indicated that the direction of defect motion was primarily tail-to-head; the motion was in the opposite direction for highly elongated cell types or epithelial cells with the cell-cell adhesion protein E-cadherin knocked out.\cite{duclos2017topological, saw2017topological, kawaguchi2017topological, blanch2018turbulent, balasubramaniam2021investigating, zhang2022, letoquin2024mechanical} By contrast, in our experiments, defects moved in either direction. To quantify these observations, we located +1/2 defects in 27 different islands across all treatments. We kept for further analysis only the defects that exhibited counter-rotating vortices (as in Fig. 1c,d) and were located at least 100 {\textmu}m away from the island boundary. Our selection criteria resulted in a total of 111 defects, which we rotated such that the tail pointed along the positive $x$ direction of the local coordinate system, with the origin being located at the head of each defect (Fig. 1e). We then computed the average $x$ component of velocity of cells along the tail of each defect in positions $0 < x < 100$ {\textmu}m, $y=0$, which we refer to as $<\!v_x\!>$. (Hereafter, we use brackets $<\!\cdot\!>$ to indicate quantities that have been averaged along the defect tails.) Interestingly, the data showed that neither tail-to-head nor head-to-tail motion was predominant (Fig. 1f). A histogram of $<\!v_x\!>$ for all defects was nearly symmetric about zero but with two well separated peaks (Fig. 1g), indicating a typical cell migration speed of a few {\textmu}m/hr with motion either head to tail or tail to head.

\subsection*{Strain Rates and Stresses near Topological Defects}

We analyzed the strain rate near the defects, focusing on the strain rate along the tail of the defect, $\dot{\varepsilon}_{xx}$. We first determined the time point at which each defect formed (which we define as $t=0$). Next, we averaged $\dot{\varepsilon}_{xx}$ for all head-to-tail and all tail-to-head moving defects at the time point immediately after defect formation ($t=0$) and 1 hr before defect formation ($t=-1$ hr). At $t=0$, the strain rates were consistent with prior reports,\cite{balasubramaniam2021investigating} namely, at the defect (near the origin), the strain rate was positive for head-to-tail moving defects and negative for tail-to-head defects (Fig. 2b,f). The difference in sign indicates that the cells elongate (for head-to-tail defects) or shorten (for tail-to-head) along the axis of each defect. Magnitudes of strain rate were on the order of 10\% hr$^{-1}$. Remarkably, for both types of +1/2 defects, the patterns of strain rate were also present at $t=-1$ hr, before defect formation (Fig. 2a,e, and Supplemental Fig. S3).

\begin{figure}[th!]
\includegraphics[width=6.5in]{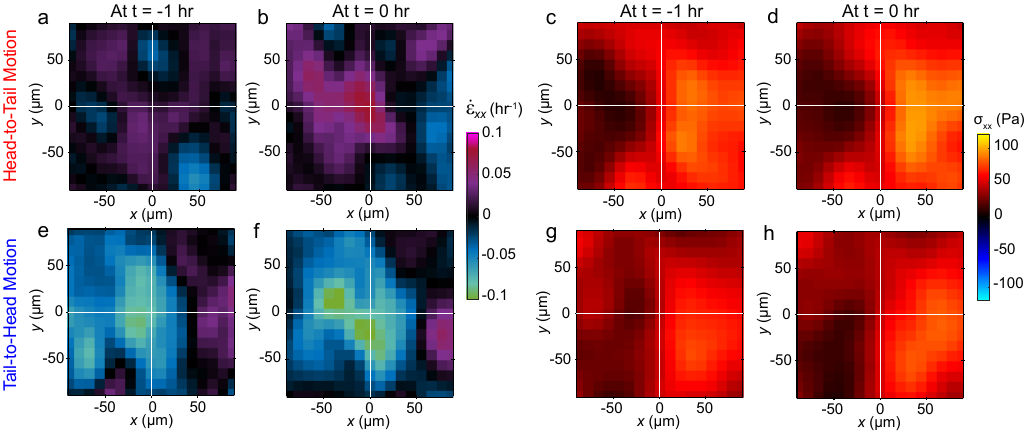}
\caption{
Strain rate and stress fields near +1/2 defects.
(a--d) For head-to-tail moving defects, the strain rate $\dot{\varepsilon}_{xx}$ and the normal stress $\sigma_{xx}$ were averaged over all 58 defects and plotted at $t=-1$ hr (1 hr before defect formation) and $t=0$ hr (the first time point after defect formation).
(e--h) For tail-to-head moving defects, $\dot{\varepsilon}_{xx}$ and $\sigma_{xx}$ were averaged over 53 defects and shown for $t=-1$ hr and $t=0$.
For all plots, the coordinate system is the same as in Fig. 1e.
}
\end{figure}

Next, we examined the stress state within the cell layer near +1/2 defects, measuring the in-plane stresses using monolayer stress microscopy.\cite{tambe2011collective, tambe2013monolayer, saraswathibhatla2020SciData} At the scale of the entire cell monolayer, there was a slight tendency for alignment between the orientations of the cells and of first principal stress, though the alignment was not perfect (Supplemental Fig. S4), as reported recently.\cite{nejad2024stress} Near +1/2 defects, past literature reported a pattern wherein the stress increased along the axis of the defect in the head-to-tail direction.\cite{saw2017topological, balasubramaniam2021investigating} As we are focused on motion along the axis of the defect (which is the $x$ direction in our chosen coordinate system), we considered the average normal stress in that direction, \textit{i.e.}, $\sigma_{xx}$, and we averaged $\sigma_{xx}$ over many defects. For both types of defects, there was a positive gradient in $\sigma_{xx}$ along the direction from head to tail (Fig. 2d,h, Supplemental Fig. S5), which is consistent with prior observations.\cite{saw2017topological, balasubramaniam2021investigating} 

The fact that the gradient in stress is the same for both tail-to-head and head-to-tail motion is puzzling, because, given the opposite direction of motion, one would expect the gradients to be opposite as well. In particular, in the case of tail-to-head motion, the cells moved against gradients in stress, \textit{i.e.}, away from regions of high tension, which is akin to moving toward regions of high pressure. Although this puzzling observation has been made before,\cite{balasubramaniam2021investigating} the underlying cause remains unclear. To investigate more deeply, we plotted $\sigma_{xx}$ an hour before defect formation, considering the same local coordinate system as of the formed defect. Similar to the strain rate, the patterns in stresses at $t=-1$ hr matched those at $t=0$ (Fig. 2c,g). These findings are distinct from prior observations in that the prior studies reported spatial patterns of strain rate and stress after the defects formed, but not before.\cite{saw2017topological, balasubramaniam2021investigating}

Next, we quantified the variability between the different defects. We calculated the average $x$ component of velocity and the gradient $\partial \sigma_{xx} / \partial x$ along the tail region of each defect, averaged the data for each defect, and plotted the results in 2D heat maps. Although there was notable variability in $<\!\partial \sigma_{xx} / \partial x\!>$, for defects moving in both the head-to-tail and tail-to-head directions, $<\!\partial \sigma_{xx} / \partial x\!>$ was positive for most defects (Supplemental Fig. S6).
We also considered that analyzing only $\sigma_{xx}$ is incomplete, as doing so neglects the other components of the stress tensor. Following prior studies, \cite{saw2017topological, balasubramaniam2021investigating} we calculated the average principal stress, $(\sigma_1+\sigma_2)/2$. Results showed similar trends, with the averaged gradient of $(\sigma_1+\sigma_2)/2$ being positive along the tail of the defect for both head-to-tail and tail-to-head defects (Supplemental Fig. S6). We also computed the gradient of stress more rigorously. As equilibrium is mathematically written as the divergence of stress, we computed the expression corresponding to equilibrium in the $x$ direction, $\partial \sigma_{xx} / \partial x + \partial\sigma_{xy} / \partial y$. Again, the sign was positive for both head-to-tail and tail-to-head moving defects (Supplemental Fig. S7).  These data confirmed that tail-to-head defects moved against gradients in stress, and, interestingly, patterns of both strain rate and stress existed before defect formation.

\subsection*{Propulsive Substrate-Cell Tractions near Topological Defects}

Motion against the stress gradient means that the stress impedes the cell motion, which is only possible if there exists another source of activity that can propel the cells against the stress gradient. A potential source of activity is the substrate-to-cell traction. We measured tractions applied by the substrate to the cells using traction force microscopy.\cite{dembo1999stresses, butler2002traction, del2007spatio, trepat2009physical}  Most models of active nematic fluids\cite{thampi2014active, duclos2018spontaneous, kumar2018tunable, shankar2018defect, angheluta2021role} assume the traction to be a viscous drag that acts as a friction. In such a case, the traction (applied by the substrate to the cell) would point exactly opposite to the direction of cell velocity. In contrast to this expectation, visual examination of a representative +1/2 tail-to-head defect showed that the fields of traction and velocity tended to align (Fig. 3a,b). The alignment was especially notable in the tail region of the defect. To study this observation in multiple defects, we averaged the tractions for all head-to-tail and tail-to-head defects and plotted the results as vector plots for $t = -1$ hr and $t = 0$ hr (Fig. 3c-f). For both cases, the traction always pointed toward the head of the defects, \textit{i.e.}, in the negative $x$ direction, which is consistent with the fact that the traction balances the stress gradient, which was in the positive $x$ direction (Fig. 2).

\begin{figure}[th!]
\includegraphics[width=6.5in]{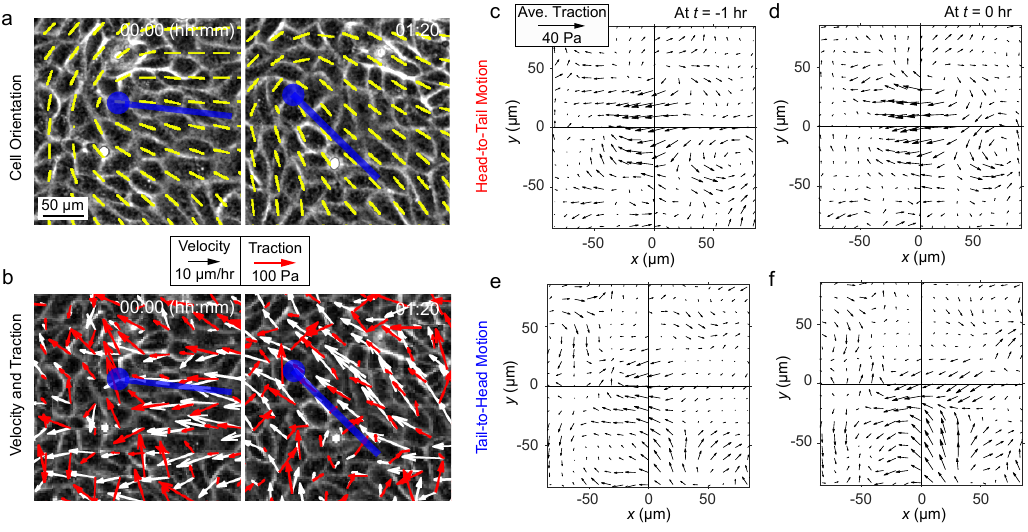}
\caption{
Traction applied by the substrate to the cells near +1/2 defects.
(a) Phase contrast images over time of a representative +1/2 tail-to-head moving defect. 
Lines indicate cell orientation. 
(b) Vector plots of velocity (white arrows) and traction (red arrows) at the same time points as in panel a. The arrows typically point in the same direction, indicating a tendency for alignment between velocity and traction for tail-to-head moving defects. 
(c, d) Plots of traction averaged over 58 head-to-tail moving defects at $t = -1$ hr and at $t = 0$ hr.
(e, f) Plots of traction averaged over 53 tail-to-head moving defects at $t = -1$ hr and at $t = 0$ hr.
}
\end{figure}

These data suggest that for tail-to-head defects, the traction acts in the same direction as the motion, meaning it is propulsive. We confirmed this observation further by plotting histograms of the angle between traction and velocity, $\theta$. For all cells across multiple monolayers, the distribution of $\theta$ was approximately uniform (Supplemental Fig. S8). A uniform distribution of $\theta$ also occurred near defect-free regions containing counter-rotating vortices (Supplemental Fig. S8). Near defects, $\theta$ showed a tendency towards 0$^\circ$ or 180$^\circ$, depending on whether the motion was tail to head or head to tail, respectively. For tail-to-head defects, the alignment between traction and velocity along the tails indicates the opposite of friction, implying that the traction propelled the cells forward (Supplemental Fig. S8). In summary, for head-to-tail moving defects, stress gradients propelled the motion, and tractions acted as a resisting friction; for tail-to-head, tractions propelled, and stress gradients resisted.

\subsection*{Energy Injection and Dissipation near Topological Defects}

The concepts of propulsion and resistance imply, respectively, injection and dissipation of energy. Following this idea, we next considered the power associated with stress and traction. To this end, we computed the power densities $P_S = \bm{\sigma}:\dot{\bm{\varepsilon}}$ and $P_T = - \bm{t}\cdot\bm{v}/h$, respectively, with $\bm{\sigma}$ and $\dot{\bm{\varepsilon}}$ being the stress and strain rate tensors, $\bm{t}$ and $\bm{v}$ being the traction and velocity vectors, $h$ being the height of the cell layer, and the symbols $:$ and $\cdot$ being the tensor and vector dot products. The signs used in our definitions of $P_S$ and $P_T$ are consistent with typical sign conventions in continuum mechanics, wherein a positive value indicates dissipation of energy, and negative indicates negative dissipation, \textit{i.e.}, injection of energy into the system by the cells. The power densities $P_S$ and $P_T$ across the full cell island fluctuated notably over space, with both positive and negative values, indicating dissipation and injection of energy, respectively (Fig. 4a,b). The spatial autocorrelation of the power densities, averaged over multiple cell islands, decayed quickly, with spatial correlation lengths of $\sim 25$ {\textmu}m, which is approximately the width of a single cell (Fig. 4c). The small correlation lengths indicate that on average, the dissipation and injection of energy is reminiscent of random noise. Interestingly, near +1/2 defects, the average power densities showed clear spatial patterns. In particular, for head-to-tail defects, both $P_S$ and $P_T$ were positive along the defect tail, indicating energy dissipation. The region of positive power densities, especially $P_T$, had a width of $\approx$100 {\textmu}m and even existed 1 hr before defect formation (Fig. 4d,e, Supplemental Fig. S9). Tail-to-head defects exhibited negative power densities, indicating energy injection, over regions of similar size (Fig. 4g,h, Supplemental Fig. S9).

\begin{figure}[b!]
\includegraphics[width=6.5in]{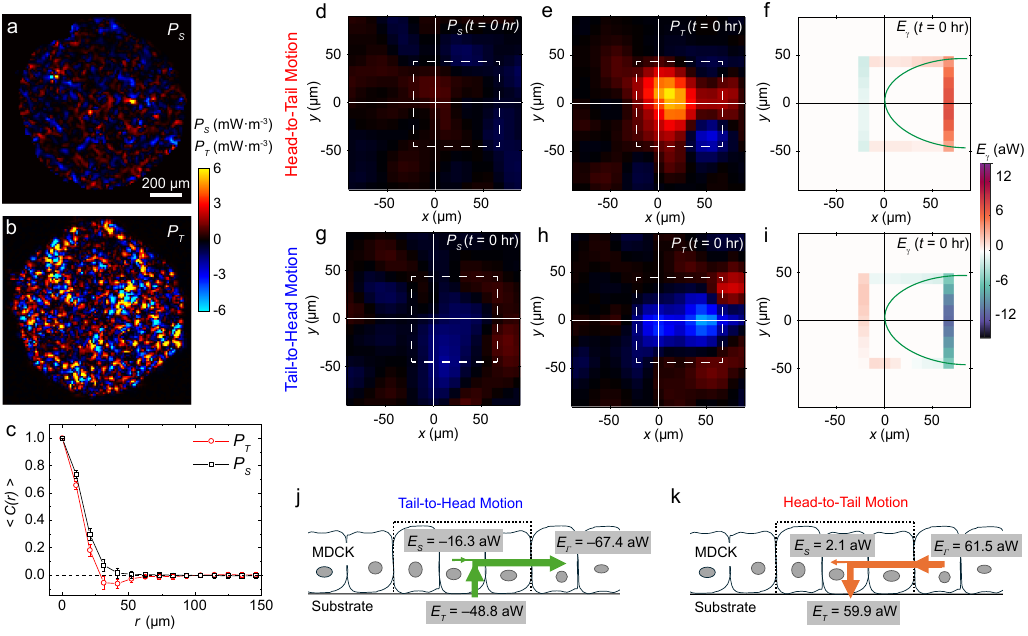}
\caption{
Power flow near +1/2 defects.
(a, b) Power densities associated with stress ($P_S$) and traction ($P_T$) for a representative 1 mm cell island. 
(c) The spatial autocorrelation of power density $C(r)$. Markers show the average over 27 different cell islands; error bars indicate the standard deviation.
(d, e) Power densities $P_S$ and $P_T$ averaged over all head-to-tail defects at $t = 0$ hr. White dotted squares represent the same region of interest (ROI) as in the next panel.
(f) Considering a 100 {\textmu}m $\times$ 100 {\textmu}m square ROI centered at (25 {\textmu}m, 0), the integral $h\int_\Gamma \bm{\sigma}\hat{\bm{n}}\cdot\bm{v} ds$ is visualized by discretizing the path $\Gamma$ into segments of length ${\Delta}s$, where ${\Delta}s$ is equal to the spatial resolution of the measured data. The colors show the quantity $E_\gamma = h \bm{\sigma}\hat{\bm{n}} \cdot \bm{v} {\Delta}s$, which indicates the flow of power across the boundary of the ROI. Red (positive) and green (negative) indicate power flowing into and out of the ROI, respectively.
(g--i) Power densities $P_S$ and $P_T$, and power flow $E_\gamma$ for tail-to-head moving defects at $t = 0$ hr.
(j, k) Schematic side view of the power flow through the ROI, for both types of defects. Dotted lines represent the ROI. Net powers associated with stress ($E_S$) traction ($E_T$), and the net power flowing through the boundaries of the ROI ($E_\Gamma$) are indicated.
}
\end{figure}

For these regions of energy dissipation and injection to exist, the energy must be transmitted to neighboring cells via cell-cell adhesions. Hence, we also quantified energy transmission using Clapeyron's theorem, which relates the net energy within a region of interest to the flow of energy across the boundaries of that region. This theorem is expressed mathematically as $\int_\Omega P_S dA + \int_\Omega P_T dA = \int_\Gamma \bm{\sigma}\hat{\bm{n}}\cdot\bm{v} ds$, where $\Omega$ defines a region of interest with perimeter $\Gamma$, and $\hat{\bm{n}}$ is the unit outward normal vector from $\Gamma$ (see Supplemental Note 1 for a derivation). To quantify energy flow near the +1/2 defects, we defined $\Omega$ to be a $100 \times 100$ {\textmu}m$^2$ region of interest (ROI) centered slightly to the right of the defect, at ($x=25$ {\textmu}m, $y=0$ {\textmu}m). To visualize the line integral $\int_\Gamma \bm{\sigma}\hat{\bm{n}} \cdot \bm{v} ds$, we discretized the path $\Gamma$ into segments of length ${\Delta}s$, where ${\Delta}s$ is equal to the spatial resolution of the measured data. We then plotted the quantity $E_\gamma = h \bm{\sigma}\hat{\bm{n}} \cdot \bm{v} {\Delta}s$, where the cell height $h$ is included to recover units of Watts. The magnitude of $E_\gamma$ was negligible on the top and bottom sides of the region of interest. Interestingly, the magnitude was largest on the right side of the defects, indicating that power flows to neighboring cells predominantly along the tail (Fig. 4f,i). For head-to-tail defects, which dissipate energy, $E_\gamma$ on the right side was positive, indicating a net flow of power inward and along the tail of the defects. The trend for tail-to-head defects was the opposite, with a net outward flow of power, also along the tail of the defects. To summarize these observations, we show schematically the flow of power in Fig. 4j,k. We indicate the three integrals in Clapeyron's theorem as follows: $E_S = h\ \int_\Omega P_S dA$; $ E_T = h\ \int_\Omega P_T dA$; and $E_\Gamma = h\ \int_\Gamma \bm{\sigma}\hat{\bm{n}}\cdot\bm{v} ds = \sum_{\Gamma} E_\gamma$. The cell layer height $h$ is included to recover units of Watts. With this new notation, Clapeyron's theorem becomes a simple balance between three quantities, $E_S+E_T = E_\Gamma$. Numerical values for $E_S$, $E_T$, and $E_\Gamma$ are shown in Fig. 4j,k, and they balance to within a couple percent, meaning errors in the measurement are small (see Supplemental Note 2 for further quantification of noise in the measurements of power). The results show that, for head-to-tail defects, power flows across the boundaries primarily along the defect tails and is dissipated primarily at the cell-substrate interface; for tail-to-head defects, power is produced primarily at the cell-substrate interface and flows outward along the defect tails. 

Following the finding that power flows primarily along the tails of the defects, and that head-to-tail and tail-to-head defects have different signs of power flow, we hypothesized that the stress supporting elements of the cytoskeleton could differ at the tails of head-to-tail and tail-to-head defects. To test this idea, we imaged cells over time and then immediately fixed the cells and fluorescently stained actin. From the time-lapse imaging, we identified head-to-tail and tail-to-head defects, and we then collected images of actin at each of the identified defects. For head-to-tail defects, actin stress fibers were commonly present near the tail of each defect, and the stress fibers aligned with the tail. For tail-to-head defects, by contrast, actin fibers tended to align perpendicular to the tail (Fig. 5a,b Supplemental Fig. S10). To quantify these observations, we manually measured the orientation of stress fibers for cells in the tail of each defect and defined an angle $\beta$ between the orientation of the stress fibers and the defect tail (Fig. 5c). For head-to-tail defects, $\beta$ tended toward zero, indicating alignment between the cells and the tail of the defects. By contrast, for tail-to-head defects, $\beta$ showed a tendency toward 90$^\circ$, indicating that the stress fibers typically aligned perpendicular to the defect tail (Fig. 5d,e). Hence, cells at the tails of head-to-tail and tail-to-head defects exhibit different cytoskeletal structures that can explain the different patterns of energy transmission. For head-to-tail defects, actin stress fibers were aligned along the defect tail and potentially transmitted energy from neighboring cells into the defect. For tail-to-head defects, actin stress fibers were aligned perpendicular to the tail, and these cells acted as a source, rather than a sink, of energy. 

\begin{figure}[th!]
\includegraphics[width=6.5in]{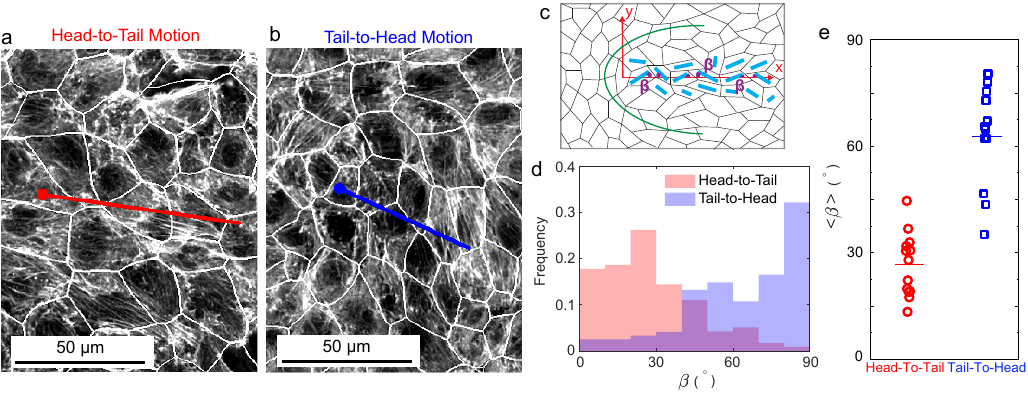}
\caption{
Orientation of stress fibers near +1/2 defects.
(a, b) Confocal images of stress fibers captured at the basal (bottom) of the cell layer for representative head-to-tail and tail-to-head defects. Cell boundaries were segmented using images collected at the apical (top) side of the cell layer and are shown overlaid on the images with white lines.
The head and tail of each defect is marked by the dot and the line, respectively.
(c) Schematic illustration showing the definition of angle $\beta$, between the orientation of stress fibers and the defect axis. 
(d) Histograms of angle $\beta$ for cells along the tails of head-to-tail and tail-to-head moving +1/2 defects.
(e) The average of $\beta$ was computed for cells along the tail of each defect and indicated by a separate marker. Solid lines indicate means. $p = 6.8 \times 10^{-5}$.
}
\end{figure}

\section*{Discussion}

We studied force and motion of epithelial cells near +1/2 topological defects. We observed that the defects moved in either the head-to-tail or tail-to-head direction, and that defects moving in both directions coexisted within the same cell monolayer. The observation of defects moving in both directions differs from prior reports in the same cell type,\cite{saw2017topological, balasubramaniam2021investigating} and, while the cause is not yet fully understood, it may be due to the fact that we studied a system at low cell density, for which the cell monolayer is likely in the fluid state.\cite{saraswathibhatla2020tractions} By observing that tractions at the cell-substrate interface are not solely dissipative as is assumed in many theories of active nematic materials, our study demonstrates how tail-to-head defects move against stress gradients, namely, they are propelled by the tractions. The resulting picture is that defect motion can be dominated either by tractions or stresses. To study more deeply what it means for stress or traction to dominate, we quantified the power associated with stresses and traction, enabling us to study the energy dynamics. For tail-to-head and head-to-tail defects, the energy flowed in opposite directions. For tail-to-head defects, which we refer to as traction dominated, energy was injected via tractions and transmitted by stresses elsewhere in the cell layer. For head-to-tail, which we refer to as stress dominated, energy was transmitted via stresses to the defect region and dissipated via tractions at the cell--substrate interface (Fig. 4j,k). Further inspection showed that actin stress fibers were aligned with the defect tails only for head-to-tail defects, which is consistent with our observation that for these defects, the energy for motion was transmitted via stresses from neighboring cells.

A recent question of interest is how +1/2 defects can move in the tail-to-head direction, which has been thought to imply extensile behavior, even through cells are contractile. Various explanations for this unexpected behavior have been proposed. One study suggested that fluctuations in each cell's shape can generate tail-to-head moving defects in a contractile system.\cite{zhang2023active} Others showed how fluctuating polar forces, such as cell-substrate tractions, can lead to extensile behavior.\cite{vafa2021fluctuations, killeen2022polar} Our data showing that traction and motion can align near defects provides new insight. The data show that traction does not act solely as a friction, as assumed by some prior models. The observation of alignment between traction and motion near +1/2 defects (Fig. 3) is surprising, because traction and velocity typically do not align, except for in special situations such as the edge of an expanding cell monolayer or next to a region on the substrate to which the cells cannot adhere.\cite{trepat2009physical, kim2013propulsion} Nevertheless, traction and velocity align near tail-to-head moving defects, meaning that for these defects tractions propel the motion, and gradients in stress act as a viscous resistance.

Hence, we show that the stresses can either propel or resist defect motion, meaning that motion of +1/2 defects is not a useful indicator of the stress state. This finding is consistent with a recent observation that the orientation of the cell body does not necessarily align with the orientation of first principal stress.\cite{nejad2024stress} A remaining question is what defines whether the energy for defect motion comes from stresses or tractions. A hint comes from our stress fiber imaging, which shows that the alignment of stress fibers differs for head-to-tail and tail-to-head moving defects. These observations are consistent with predictions of a recent theory that proposed the existence of  two distinct orientation fields in the epithelial cell layer, one for the shape of the cell body and the second for the stress-generating actomyosin fibers.\cite{nejad2025cellular}

Our quantification of power provides a useful way to study the joint effect of stresses and tractions, and the data reveal notable fluctuations of power density over space. Interestingly, we found non-negligible spatial correlations in power at +1/2 defects, with the sign of power density corresponding to the direction of motion of the defect. Although the cause for these spatial correlations is not yet clear, spatial correlations in tractions have been observed before in epithelial cell monolayers\cite{saraswathibhatla2021, vazquez2022effect} and in monolayers of bacteria near topological defects.\cite{han2025local} Given that, in our experiments, the spatial correlations in traction and stress were present at $t=-1$ hr (\textit{i.e.}, before defect formation), it is reasonable to expect that they are important for defect formation.

An important observation in this study is that patterns of strain rate, stress, traction, and power density exist in the cell monolayer even an hour before defect formation. Most notable is that the signs of strain rate and power density differ for head-to-tail and tail-to-head defects. Given these observations, it is likely that the direction of defect motion is determined before the defect forms. In the case of tail-to-head moving defects, before the defect forms, the cells apply traction in coordination with their neighbors (Fig. 3e,f), leading to collective motion and a negative strain rate (Fig. 2e,f). In turn, these cells shrink in the $x$ direction, causing them to acquire a long axis along the $y$ direction, which forms a +1/2 defect that is moving in the tail-to-head direction. In the case of the head-to-tail moving defects, locally high tensile stresses may pull on a group of cells, causing them to increase in size (positive strain rate, Fig. 2a,b). These cells resist motion with a net dissipation of energy (Fig. 4d,e), but are pulled along by energy transmitted through stress fibers along the tail of the defect (Fig. 4f, and Fig. 5d,e). 

Together, our data suggest that the defects form as a result of the collective motion of a local group of cells. Nematic order would not need to exist initially---the nematic order could emerge from the collective motion. The notion that nematic order and defects result from patterns in cell motion and strain rate differs from current understanding---it is commonly assumed that a tendency for nematic order defines the local cell orientation, which in turn defines the orientation of active stresses and the resulting cell motion. Our findings change the focus, suggesting that cell motion may cause formation of nematic order and defects. This idea is conceptually similar to recent models that showed how a balance of intercellular forces can cause cells to elongate, in turn leading to nematic order and topological defects,\cite{zhang2023active, chiang2024} though a difference is that our experiments also show spatially correlated patterns of cell tractions and stresses leading to formation of topological defects. In future work, it will be interesting to study what causes these spatial patterns of cell forces.

\section*{Methods}

\subsection*{Cell Culture}

MDCK type II cells (acquired from MilliporeSigma) and MDCK type II cells expressing green fluorescent protein attached to a nuclear localization signal (a gift from Professor David Weitz, Harvard University) were maintained in 1 mg/ml glucose Dulbecco's modified Eagle's medium (10-014-CV, Corning Inc.) supplemented with 10\% fetal bovine serum (Corning). The cells were maintained in an incubator at 37$^\circ$C and 5\% CO$_2$.

\subsection*{Sample Preparation}

Experiments were performed with cells seeded in micropatterned islands of 1 mm diameter on polyacrylamide substrates of Young's modulus 6 kPa having fluorescent particles localized at the top of the gel for traction force microscopy (TFM). Glass bottom dishes (Cellvis) were activated with plasma treatment and coated with 3-(Trimethoxysilyl) propyl methacrylate (MilliporeSigma). A freshly made polyacrylamide solution (5.5\% w/v acrylamide, 0.2\% w/v bisacrylamide, Biorad Laboratories) was mixed with 0.03\% w/v fluorescent particles (diameter 0.5 {\textmu}m, carboxylate modified, Life Technologies) and 50 {\textmu}l of gel and bead solution was deposited on the activated glass surface. A coverslip was placed on top of the polyacrylamide solution before polymerization, and the gel was centrifuged upside down to localize the fluorescent particles to the top. The final thickness of the gels was approximately 100 {\textmu}m. For micropatterning of 1 mm cell islands, polydimethysiloxane (PDMS) (Sylgard 184) was cured into a 200 {\textmu}m thick sheet, and 1 mm holes were cut with a biopsy punch to create PDMS masks. The PDMS masks were sterilized with 70\% ethanol and submerged in 2\% Pluronic F-127 (Sigma-Aldrich) overnight at room temperature. A treated PDMS mask was placed on top of each polyacrylamide gel, and the gels were functionalized using sulfo-SANPAH (100 {\textmu}g/ml, Pierce Biotechnology) to covalently crosslink rat tail collagen I (Corning). 1 ml of 0.1 mg/ml collagen was added to each gel and the gels were incubated overnight at 4$^\circ$C. Approximately $0.5\times10^6$ cells (in 0.5 ml of medium) were put on top of each gel for 2 hr to adhere, followed by removal of the PDMS mask and replacement of the cell culture medium. The seeded cells were allowed to spread to confluence in each island for 24 hr in medium containing 2\% fetal bovine serum. The experiments were performed at a low density of 2400 cells/mm$^2$, which resulted in relatively long aspect ratios and greater cell motion than occurs at higher densities.\cite{saraswathibhatla2020tractions} The low density also prevented extrusion of cells out of the monolayer near +1/2 defects, which has been reported in prior studies.\cite{saw2017topological, copenhagen2021topological, han2025local} Extrusion events were not desirable in this study, as they could have confounding effects on the cell motion.

\subsection*{Microscopy}

Time lapse imaging of cells and fluorescent particles was performed using an Eclipse Ti microscope (Nikon Instruments, Melville, NY) with a $10\times$ numerical aperture 0.3 objective or a $20\times$ numerical aperture 0.5 objective (Nikon) and an Orca Flash 4.0 camera (Hamamatsu, Bridgewater, NJ) controlled by Elements Ar software (Nikon). Cells were maintained at 37$^\circ$C and 5\% CO$_2$ using a custom-built cage incubator. Images were collected every 15 min. Immediately before imaging, some cell islands were treated with inhibitors and activators of actomyosin contraction, namely, CN03 (2 {\textmu}g/mL Cytoskeleton, Inc.) and CN02 (40 ng/ml, Cytoskeleton). After imaging, the cells were removed by incubating in 0.05\% trypsin for 1 hr, and images of the fluorescent particles in the substrate were captured for a traction-free reference state to compute cell-substrate tractions.

For imaging of stress fibers, cells in the 1 mm cell islands and their green fluorescent protein-labeled nuclei were imaged by phase contrast and fluorescence microscopy using the Eclipse Ti microscope for 8 hr and then immediately the cells were fixed in a 4\% paraformaldehyde solution for 20 min and permeabilized using 0.1\% Triton X-100. The last frame of phase contrast images was used to identify head-to-tail and tail-to-head moving +1/2 defects at the end of the 8 hr experiment (Supplemental Fig. S10a).  Identified defects were plotted along with the fluorescent images of their nuclei (Supplemental Fig. S10b). Next, the fixed cells were stained for actin imaging using ActinRed 555 ReadyProbe Reagent (Invitrogen, catalog number R37112) according to manufacturer instructions. Confocal images of nuclei (Supplemental Fig. S10c) and of actin (Supplemental Fig. S10d,e) were collected on a Nikon A1R confocal microscope with a water immersion $40\times$ numerical aperture 1.15 objective (Nikon) using NIS-Elements Ar software. The images of the nuclei were used as a guide to match the confocal images to the phase contrast images acquired during the time lapse, as described in Supplemental Fig. S10. 

\subsection*{Quantification of Cell Velocity and Orientation}

The cell velocity field $\bm{v}$ was calculated from phase contrast images of cells by using Fast Iterative Digital Image Correlation (FIDIC) \cite{bar2015fast} using $32 \times 32$ pixel ($21 \times 21$ {\textmu}m$^2$) subsets centered on a grid with spacing of 8 pixels (5.2 {\textmu}m) and dividing by the imaging time (15 min). The vorticity was calculated from the velocity $\bm{v}$ according to $\partial v_{y}/\partial x-\partial v_{x}/\partial y$. The strain rate tensor $\dot{\bm{\varepsilon}}$ was calculated according to $\dot{\bm{\varepsilon}} = [\nabla \bm{v} + (\nabla \bm{v})^T ]/2$, with $(\cdot)^T$ indicating the transpose. Numerical derivatives were computed by the central difference method. Defect identification was similar to methods reported previously \cite{saw2017topological, copenhagen2021topological} and based on the local orientation of the cells, which was determined by analysis of phase contrast images. Phase contrast images were analyzed with the ImageJ plugin OrientationJ using a window size of 16 pix (10.4 {\textmu}m) and the ``Cubic Spline'' option for computing the gradient. Orientation fields were plotted for manual selection of defects. To rule out that candidate defects identified were due to errors in quantifying the cell orientations, we used the vorticity as a verification criterion, and kept only +1/2 topological defects having two counter-rotating vortices on either side of the tail (as in Fig. 1c,d) for further analysis. Given that the cells tend to align with boundaries of the cell island and the cell motion near boundaries is restricted to the azimuthal direction, defects within 100 {\textmu}m of the boundaries of the cell islands were excluded from the analysis. 

\subsection*{Quantification of Tractions and Stresses}
 
For traction force microscopy, substrate displacements were computed by applying FIDIC to images of the fluorescent particles using $32 \times 32$ pixel (21 $\times$ 21 {\textmu}m$^2$) subsets centered on a grid with spacing of 8 pixels (5.2 {\textmu}m). Traction was calculated with unconstrained Fourier-transform traction microscopy accounting for substrates of finite thickness.\cite{butler2002traction, del2007spatio, trepat2009physical} In this manuscript, the tractions that are reported are the traction applied by the substrate to the cells. The stresses within the cell layer were calculated with monolayer stress microscopy \cite{tambe2011collective, tambe2013monolayer} (with freely available software \cite{saraswathibhatla2020SciData}) using a cell layer height of $h=5$ {\textmu}m. Briefly, monolayer stress microscopy computes the three components of the in-plane stress tensor from the traction data. Two of the three equations required are force equilibrium in the two in-plane directions, which make no assumptions about the rheology of the cell layer or the stress state. The third equation comes from the compatibility equation of continuum mechanics, which enforces smoothness over space. Applying this third equation assumes a linear, passive relationship between stress and strain rate. The implementation is equivalent for elastic or viscous rheology and depends only on a single dimensionless constant, which is equal to the ratio of shear to bulk modulus (for elastic) or viscosity (for viscous) and was chosen to be 0.54 for this system. Prior work has shown dependence on this constant to be negligible.\cite{tambe2013monolayer} Given that the active component of the stresses is not accounted for in the compatibility equation used in monolayer stress microscopy, it is important to consider potential errors. A theoretical model for collective migration studied these errors,\cite{zimmermann2014intercellular} with results suggesting that if the magnitude of traction is sufficiently large (\textit{e.g}., at least 1/3 of the magnitude of stress), errors in the stresses calculated by monolayer stress microscopy become negligible. As shown in Supplementary Fig. S1, this requirement is met. Our experiments used cell islands, which is important for monolayer stress microscopy, as it allows for all boundaries to be imaged such that a stress-free boundary condition can be used. Three additional Dirichlet boundary conditions are required to stabilize the computation. For a cell island for which the tractions are in a state of equilibrium, these boundary conditions do not affect the stress state, which we verified, as described previously. \cite{saraswathibhatla2020SciData}

\subsection*{Spatial Autocorrelation}

The spatial autocorrelation of power density $P$ was calculated according to
\[
C(r) = \frac{\sum \bar{P}(\bm{r}^\prime)\bar{P}(\bm{r}^\prime - \bm{r})}{\sum \bar{P}(\bm{r}^\prime)^2},
\]
with $\bar{P}=P-P_m$ and $P_m$ being the average of $P$ over space, $\bm{r}$ and $\bm{r}^\prime$ indicating positions in space, and the scalar $r$ being the magnitude of vector $\bm{r}$. The sum $\sum$ was taken over all positions in the cell island. The correlation length was defined as the distance over which $C(r)$ decreased to 0.2.

\subsection*{Analysis of Confocal Images of Actin}

For analyzing confocal images, the upper (apical) $z$-planes of actin images were used to identify the cell boundaries using Cellpose version 3.0.7 for segmentation.\cite{stringer2021cellpose} The cell boundaries were fed to OrientationJ using a window size of 64 pix (19.84 {\textmu}m) and the ``Cubic Spline'' option for computing the gradient. The orientation fields were overlaid with the cell boundaries to confirm accuracy of the segmentation (Supplemental Fig. S10d). Then the lower (basal) $z$-planes of the confocal actin image stacks were overlaid with the cell boundaries and the +1/2 defect to quantify manually the average orientation of stress fibers for each fixed cell on the defect axis (Supplemental Fig. S10e). 

\subsection*{Statistical Analysis}

Comparisons between two groups were performed with the rank sum statistical test. Comparisons between more than two groups were performed using the Kruskal-Wallis test with Bonferroni correction for multiple comparisons.

\section*{Data and Code Availability}

A partial data set of this study, including unprocessed images of cells and fluorescent particles, and processed data, is available at https://doi.org/10.6084/m9.figshare.28774310 (ref. \cite{berafigshare2025defects}). The full data set will be shared by the lead contact upon request.  Code used to compute cell velocities, tractions, and stresses is available from https://github.com/jknotbohm/FIDIC and https://github.com/jknotbohm/Cell-Traction-Stress.

\section*{Author Contributions}
P.K.B., M.M., J.Z., and J.N. designed the experiments. M.M. and J.Z. performed the experiments. P.K.B., J.Z., and J.N. analyzed data. P.K.B. and J.N. wrote the manuscript.

\section*{Declaration of Interests}
The authors declare no competing interests.

\section*{Acknowledgments}

We thank Christian Franck for use of the confocal microscope. We thank Fridtjof Brauns for insightful discussions held at the Kavli Institute for Theoretical Physics (KITP), which is supported by NSF grant number PHY-2309135 and Gordon and Betty Moore Foundation Grant number 2919.02. This work was supported by the University of Wisconsin-Madison Office of the Vice Chancellor for Research and Graduate Education with funding from the Wisconsin Alumni Research Foundation, National Science Foundation grant number CMMI-2205141, and National Institutes of Health grant number R35GM151171.

\section*{Supplemental Information}

\section*{Supplemental Note 1: Derivation of Clapeyron's Theorem}

Here, we show the derivation for Clapeyron's Theorem, which represents the balance between the three terms plotted in Fig. 4. We begin by integrating $\bm{\sigma} : \bm{\dot{\varepsilon}}$ over some region of interest $\Omega$, noting that the strain rate $\bm{\dot{\varepsilon}}$ is related to the gradient of velocity $\bm{v}$, enabling us to write
\begin{equation}
\int_\Omega \bm{\sigma} : \bm{\dot{\varepsilon}} dA = \int_\Omega \bm{\sigma} : \frac{1}{2}\left[ \nabla\bm{v} + (\nabla\bm{v})^T \right] dA = \int_\Omega \bm{\sigma} :  (\nabla\bm{v}) dA,
\label{power1}
\end{equation}
where the last expression results from symmetry of the stress tensor. The right hand side of Eq. \ref{power1} can be rewritten to give
\begin{equation}
\int_\Omega \bm{\sigma} : \bm{\dot{\varepsilon}} dA = \int_\Omega \nabla \cdot (\bm{\sigma} \bm{v}) dA - \int_\Omega (\nabla \cdot \bm{\sigma}) \cdot \bm{v} dA.
\label{power2}
\end{equation}
The right hand side of Eq. \ref{power2} can be simplified by using the divergence theorem and noting that, by equilibrium, $\nabla \cdot \bm{\sigma} = - \bm{t}/h$ to give
\begin{equation}
\int_\Omega \bm{\sigma} : \bm{\dot{\varepsilon}} dA = \int_\Gamma \bm{\sigma} \hat{\bm{n}} \cdot \bm{v} ds + \int_\Omega \frac{\bm{t}}{h} \cdot \bm{v} dA,
\label{clapeyron}
\end{equation}
where $\Gamma$ is the path enclosing region $\Omega$, and $\hat{\bm{n}}$ is the unit outward normal vector from $\Gamma$.
Note that no constitutive relationship is assumed in deriving Eq. \ref{clapeyron}. This equation is Clapeyron's Theorem, and it describes the flow of power into and out of the region of interest, with the first term on the right hand side representing power flowing through the boundaries and transmitted through stresses and the second term representing power generated or dissipated by tractions at the cell-substrate interface.
Finally, as described in the main text, we define $P_S = \bm{\sigma} : \bm{\dot{\varepsilon}}$ and $P_T = -\bm{t} \cdot \bm{v}/h$, enabling us to rewrite Clapeyron's Theorem as
\begin{equation}
\int_\Omega P_S dA + \int_\Omega P_T dA = \int_\Gamma \bm{\sigma} \hat{\bm{n}} \cdot \bm{v} ds.
\end{equation}
The units of each integral above are the power per length, where the length is the height of the cell layer. Typically, the integrals above are each described as of a power; the reason for the discrepancy is that the integrals are written using a two-dimensional formulation, whereas our manuscript reports the stress $\bm{\sigma}$ as a three-dimensional quantity with units of force per length squared. To report values of power in this manuscript, we multiplied each integral by the height of the cell layer, $h$.

\section*{Supplemental Note 2: Estimation of Noise Floor in Power Density Measurements}

Given that the powers $E_S$, $E_T$, and $E_\Gamma$ do not balance perfectly in Fig. 4j,k, here we analyze in detail the noise in the measurements of power. We consider the full cell island, for which $\bm{\sigma}\hat{\bm{n}} = 0$, meaning $E_\Gamma=0$ and then we expect $E_S + E_T = 0$. For this analysis, we choose to use units of power density (power/volume), as any error in the measurement can be directly compared to the values of power density shown in Fig. 4a,b. The quantities $E_S$ and $E_T$, calculated for the full cell island and normalized by the island volume, are shown in Fig. S11. As can be seen, their sum is in the range of 0.01--0.03 mW/m$^3$, which is nearly a factor of 100 smaller than the typical values of power density shown in Fig. 4a,b, which are on the order of 1 mW/m$^3$. Hence, the powers balance, and the level of noise in the data is negligible.

\vspace{11pt}
 
\section*{Supplemental Figures}

% FIGURE S1
\begin{center}
\includegraphics[width=6.5in]{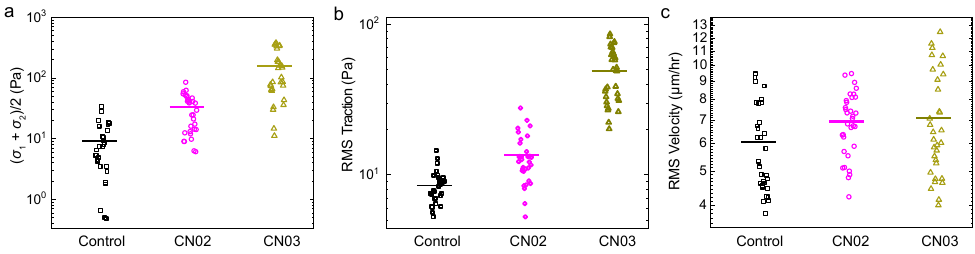}
\end{center}
\vspace{-11pt plus 1pt minus 2pt}
\textbf{Supplemental Figure S1.} 
Effects of CN02 and CN03 treatments on stress, traction, and motion.
(a) Average of principal stress $(\sigma_{1}+\sigma_{2})/2$. Each marker represents the mean of $(\sigma_{1}+\sigma_{2})/2$ over all cells in an independent cell island. 
(b,c) Root-mean-square (RMS) of traction (b) and velocity (c). Each marker represents RMS over all cells in an independent cell island.
The $p$-values for comparisons between different groups in panel (a) are $p_{control-CN02} = 3.8\times10^{-4}$, $p_{CN02-CN03} = 1.4\times10^{-4}$, $p_{control-CN03} = 7.9\times10^{-15}$, in panel (b) are $p_{control-CN02} = 0.006$, $p_{CN02-CN03} = 4.0\times10^{-7}$, $p_{control-CN03} = 1.4\times10^{-16}$, and in panel (c) are $p_{control-CN02} = 0.07$, $p_{CN02-CN03} = 1$, $p_{control-CN03} = 0.23$.

% FIGURE S2
\begin{center}
\includegraphics[width=5in]{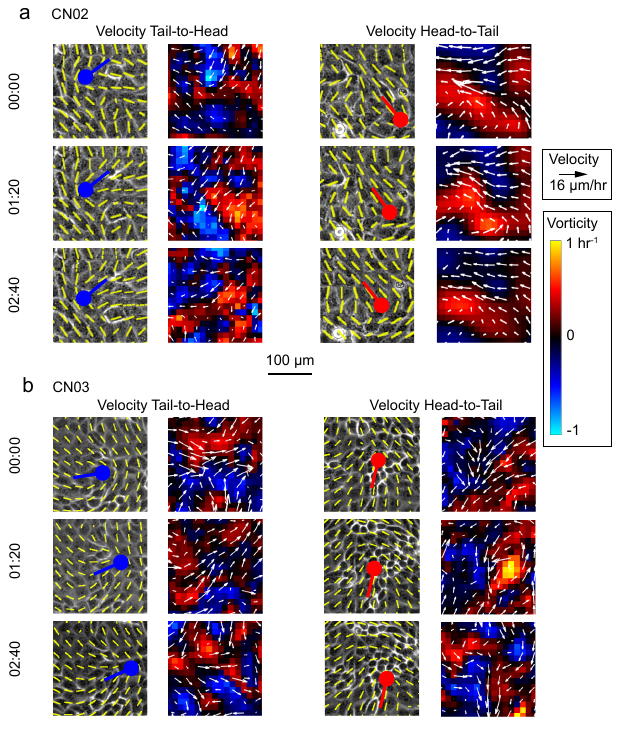}
\end{center}
\vspace{-11pt plus 1pt minus 2pt}
\textbf{Supplemental Figure S2.} 
Additional examples of velocity field near +1/2 defects. Defects moving in the tail-to-head and head-to-tail directions for cells treated with CN02 (a) and CN03 (b). Yellow lines show cell orientations, white lines show velocity, and colors show vorticity.

% FIGURE S3
\begin{center}
\includegraphics[width=6.5in]{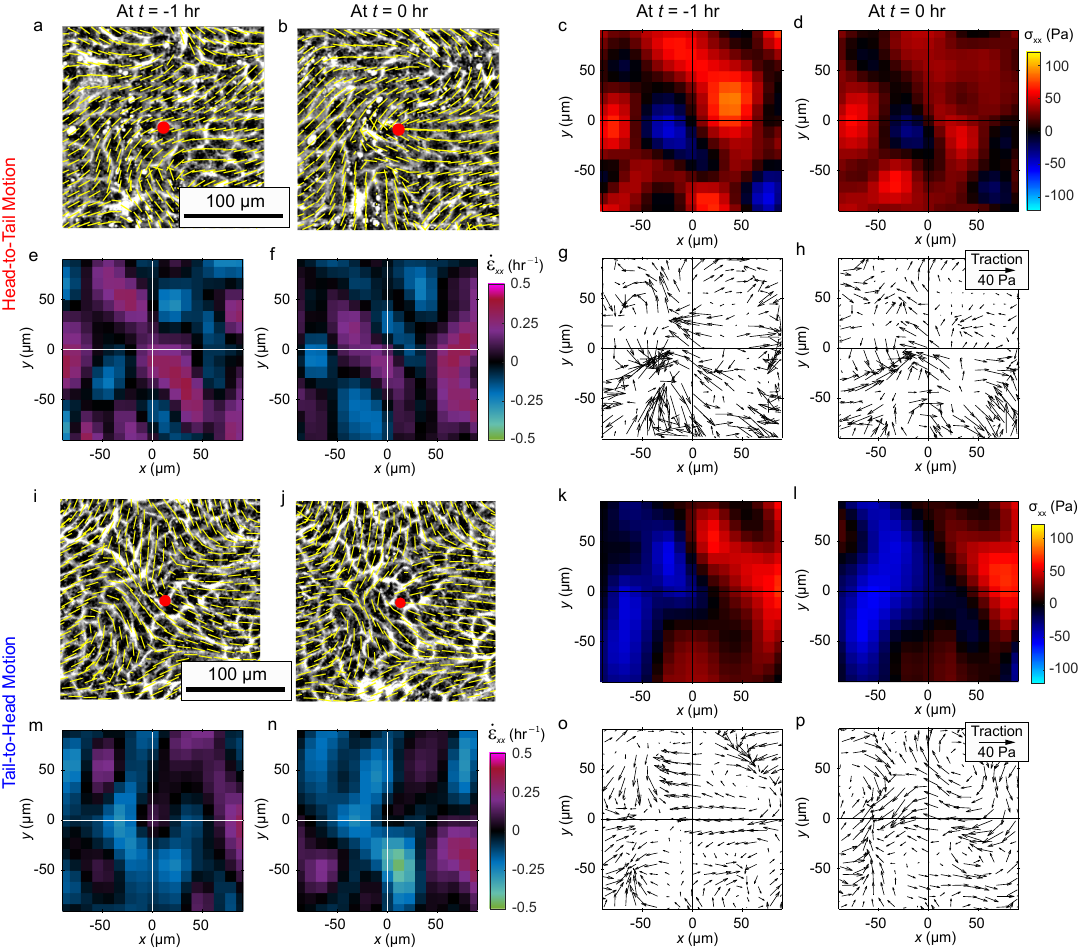}
\end{center}
\vspace{-11pt plus 1pt minus 2pt}
\textbf{Supplemental Figure S3.} 
Plots of strain rate $\dot{\varepsilon}_{xx}$, stress $\sigma_{xx}$, and traction for a single head-to-tail defect and a single tail-to-head defect. Time points shown are 1 hr before defect formation ($t=-1$ hr) and immediately after defect formation ($t=0$).

\newpage
% FIGURE S4
\begin{center}
\includegraphics[width=6.5in]{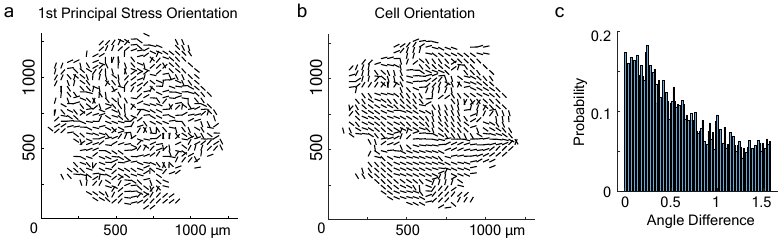}
\end{center}
\vspace{-11pt plus 1pt minus 2pt}
\textbf{Supplemental Figure S4.} 
Orientations of first principal stress and cell body.  
(a) Orientation of first principal stress in a representative cell island. 
(b) Cell orientations for the same cell island as shown in panel a. 
(c) Histogram of angle difference between the first principal orientation and cell orientation in the 1 mm cell island. 

% FIGURE S5
\begin{center}
\includegraphics[width=4.1in]{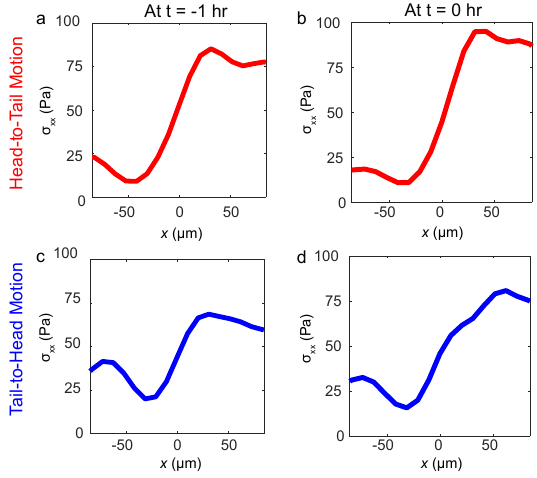}
\end{center}
\vspace{-14pt plus 1pt minus 2pt}
\textbf{Supplemental Figure S5.} 
Gradient of stress along the tail of +1/2 defects. (a,b) For head-to-tail moving defects, the average of $\sigma_{xx}$ is plotted against $x$ position along the defect tail (\textit{i.e.}, for $y=0$) at $t = -1$ hr (a) and at $t = 0$ (b).
(c,d) Similar plots for tail-to-head moving defects. Near the defect head (\textit{e.g.}, $0 \leq x\leq 50$ {\textmu}m), the gradient of $\sigma_{xx}$ is positive for both cases.

% FIGURE S6
\begin{center}
\includegraphics[width=6.5in]{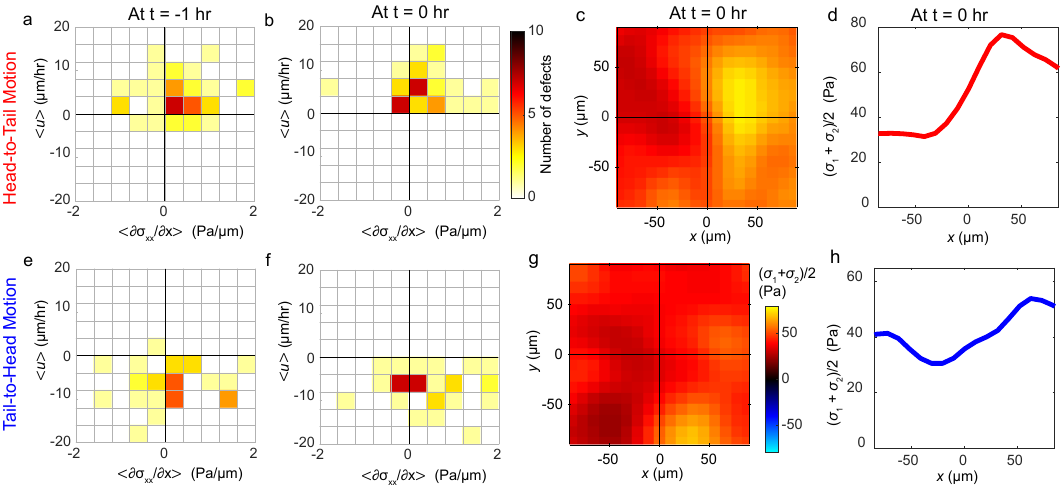}
\end{center}
\vspace{-11pt plus 1pt minus 2pt}
\textbf{Supplemental Figure S6.} 
The gradient of stress and the average principal stress near +1/2 defects. 
(a,b) Considering head-to-tail defects, heat maps of number distribution in the [$<\!v_x\!>$, $<\!\partial \sigma_{xx} / \partial x\!>$] plane at $t = -1$ hr and at $t = 0$ hr, respectively. 
Data were averaged along each defect tail ($0 \leq x\leq 50$ {\textmu}m, $y = 0$). 
The plots indicate notable heterogeneity with the magnitude of $<\!v_x\!>$ varying from 0 to 20 {\textmu}m/min and the magnitude of $<\!\partial \sigma_{xx} / \partial x\!>$ ranging from 0 to 2 Pa/{\textmu}m.
(c,d) For head-to-tail defects, map of average principal stress $(\sigma_{1}+\sigma_{2})/2$, and its variation along the tail (\textit{i.e.}, along $y=0$) at $t = 0$ hr. 
(e-h) Similar plots for tail-to-head moving defects. 

\newpage
% FIGURE S7
\begin{center}
\includegraphics[width=5.5in]{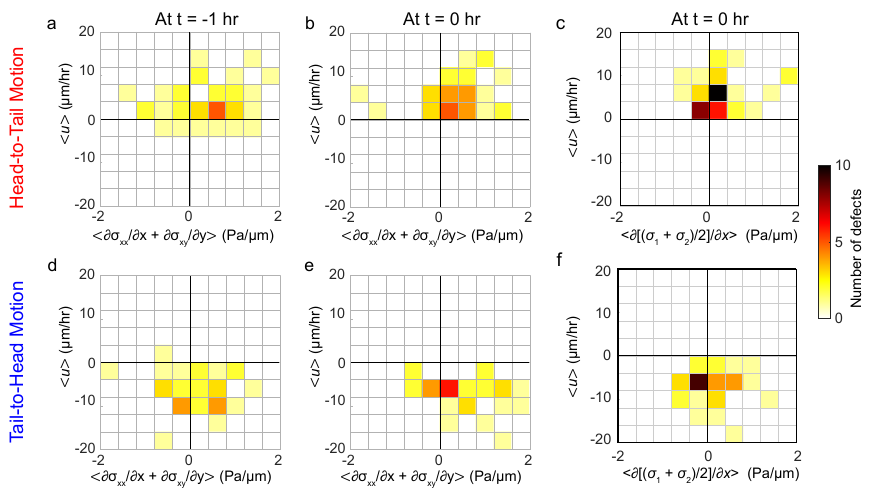}
\end{center}
\vspace{-18pt plus 1pt minus 2pt}
\textbf{Supplemental Figure S7.} 
The gradient of stress via the divergence near +1/2 defects. 
(a,b) Considering head-to-tail defects, heat maps of number distribution in the [$<\!v_x\!>$, $<\!\partial \sigma_{xx} / \partial x + \partial \sigma_{xy} / \partial y \!>$] plane at $t = -1$ hr and at $t = 0$ hr respectively. 
Data were averaged along each defect tail ($0 \leq x\leq 50$ {\textmu}m, $y = 0$). 
Here, $\partial \sigma_{xx} / \partial x + \partial \sigma_{xy} / \partial y$ is the $x$ component of the stress equilibrium expression $\nabla \cdot \bm{\sigma}$, with $\bm{\sigma}$ being the stress tensor. 
(c) For head-to-tail defects, the number distribution heat map in the  $<\!v_x\!>$ and $<\!\partial [(\sigma_{1}+\sigma_{2})/2] / \partial x\!>$ plane is shown at $t = 0$ hr.
(d-f) Similar plots for tail-to-head moving defects. 

% FIGURE S8
\begin{center}
\includegraphics[width=6.5in]{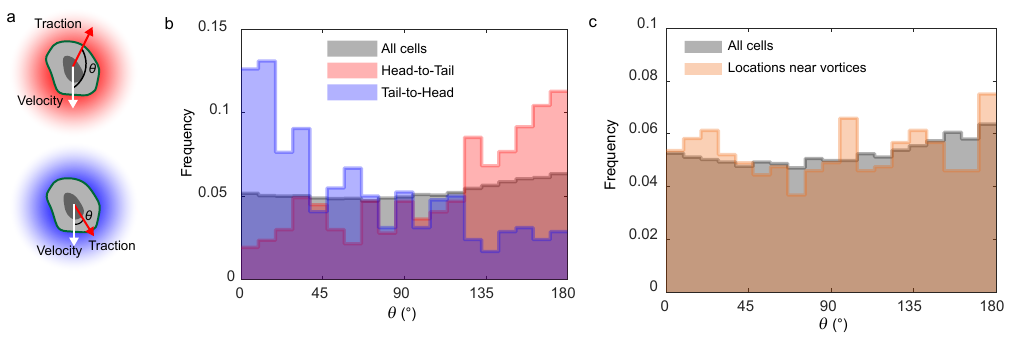}
\end{center}
\vspace{-22pt plus 1pt minus 2pt}
\textbf{Supplemental Figure S8.} 
Angle $\theta$ between traction and velocity near +1/2 defects, and near defect-free counter-rotating vortices. 
(a) The schematic illustrations show possible angles ($\theta$) between traction (applied by the substrate onto the cells) and cell velocity. If $\theta > 90^\circ$ (red shading), traction acts as a friction; if $\theta < 90^\circ$ (blue shading), traction is propulsive. 
(b) Histograms of $\theta$ at $t = 0$ hr, considering all cells in 27 islands (gray) and cells located on the defect tail in the range $0 \leq x\leq 100$ {\textmu}m for head-to-tail (red) and tail-to-head (blue) defects, respectively.
(c) Angle $\theta$ was also measured in 25 defect-free regions (length 100 {\textmu}m, width 20 {\textmu}m) having counter-rotating vortices. The distribution (orange) is nearly flat, similar to the histogram of $\theta$ for all cells (gray), which rules out the alternate explanation that counter-rotating vortices themselves caused the tendency toward 0 and 180$^\circ$.

% FIGURE S9
\begin{center}
\includegraphics[width=6.5in]{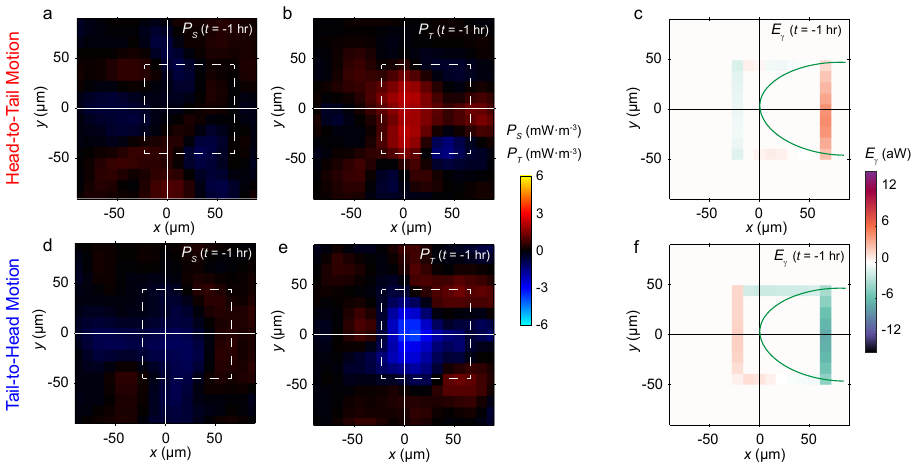}
\end{center}
\vspace{-14pt plus 1pt minus 2pt}
\textbf{Supplemental Figure S9.} 
Power flow at $t = -1$ hr near +1/2 defects.
(a, b, c) At $t = -1$ hr near head-to-tail defects, power densities $P_S$, $P_T$, and the flow of power from the side cells $E_\gamma$ are shown for the ROI.
(d--f) Similar plots are shown for tail-to-head moving defects at $t = -1$ hr.

% FIGURE S10
\begin{center}
\includegraphics[width=5.5in]{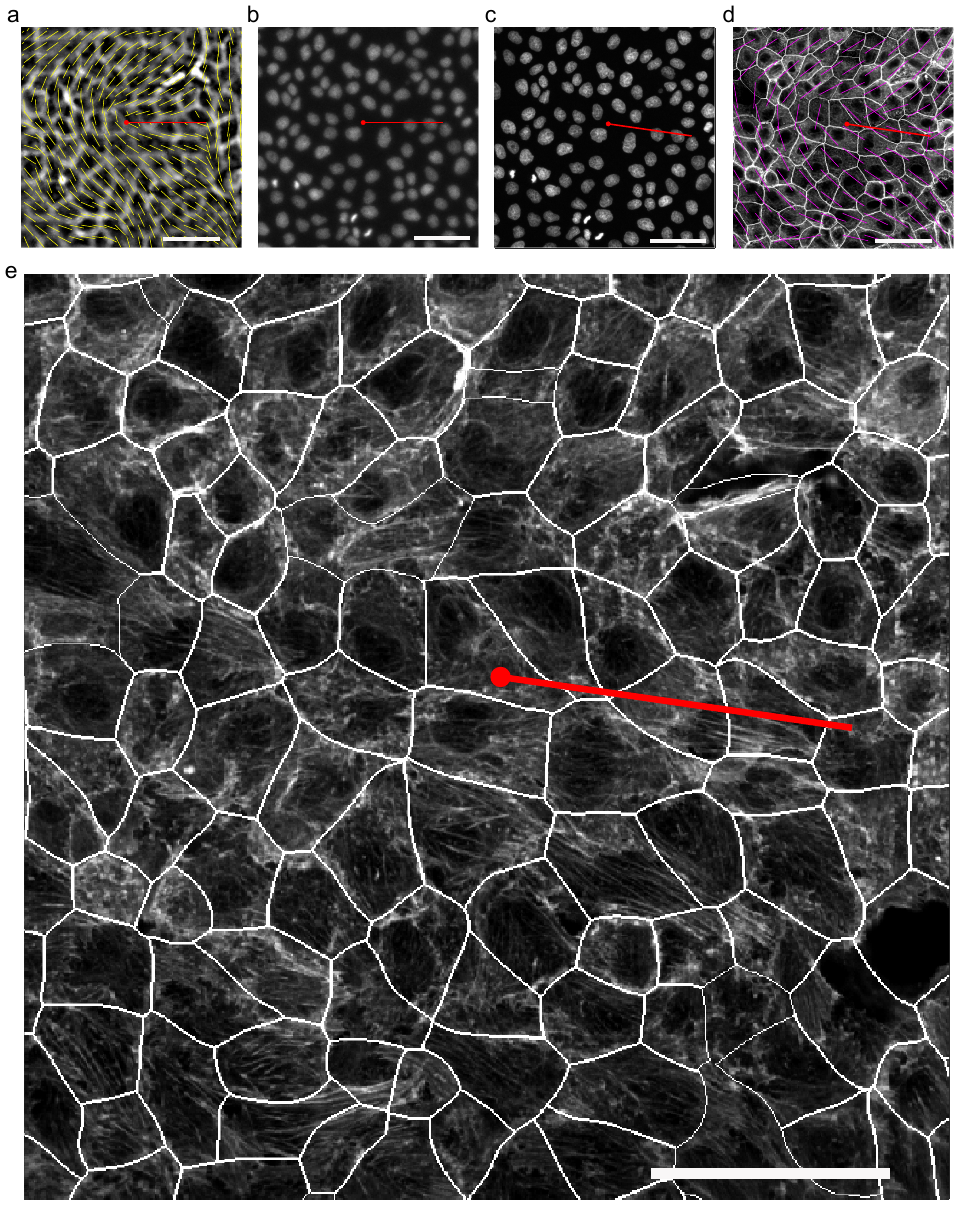}
\end{center}
\vspace{-14pt plus 1pt minus 2pt}
\textbf{Supplemental Figure S10.} 
Figure showing how stress fiber orientations near +1/2 defects were identified.  
(a, b) Phase contrast image and image of fluorescent nuclei near a representative +1/2 defect before fixing cells. Yellow lines indicate cell orientation. The identified defect is indicated in red. 
(c) Confocal image of nuclei after fixing cells, which was used to identify the location of the defect of interest.
(d) Confocal image of actin at the apical (top) side of the cell layer. White lines show segmented cell boundaries. Cell orientations were determined by applying OrientationJ to the the images of apical actin (shown in magenta) to verify the presence of a +1/2 defect.
(e) Segmented cell boundaries from the apical images (shown in panel d) are overlaid on an image of the actin stress fibers on the basal (bottom) side of the cell monolayer. To quantify angle $\beta$, the orientations of stress fibers were determined manually for each cell along the defect tail.
Scale bars represent 50 {\textmu}m.

% FIGURE S11
\begin{center}
\includegraphics[width=6.5in]{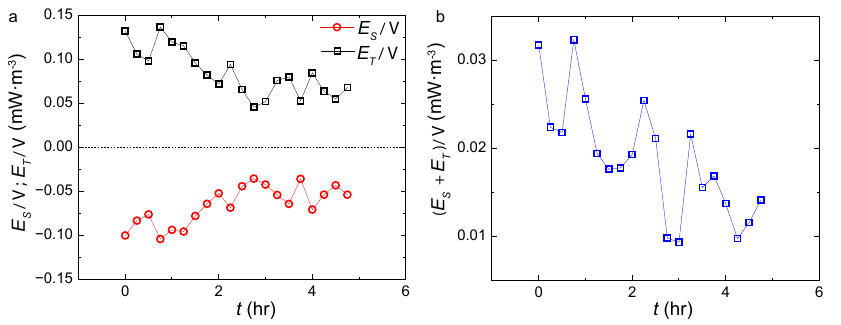}
\end{center}
\vspace{-14pt plus 1pt minus 2pt}
\textbf{Supplemental Figure S11.}
Verification of global energy balance for a representative island. For a full cell island, $\bm{\sigma}\hat{\bm{n}}=0$, meaning $E_\Gamma=0$. (a) Power densities $E_S/V$ and $E_T/V$ are plotted over time for a representative full island, where $V$ is the volume of the cell island. (b) The sum $(E_S+ E_T)/V$ is shown over time. Differences from zero are due to experimental noise, which is small compared to the values power density shown in Fig. 4a,b.

\section*{Supplemental Video Captions}

\textbf{Video 1. Representative 1 mm diameter MDCK cell island.} The red box indicates a +1/2 defect moving in the head-to-tail direction, and the blue box indicates a +1/2 defect moving in the tail-to-head direction. The experiment time is indicated in the upper left corner.

\textbf{Video 2. Zoomed in video of the red region in Video 1.} The red dot indicates the head of the head-to-tail moving +1/2 defect. The arrow on the defect tail indicates the movement of the defect.

\textbf{Video 3. Zoomed in video of the blue region in Video 1.} The blue dot indicates the head of the tail-to-head moving +1/2 defect. The arrow on the defect tail indicates the movement of the defect.

% REFERENCES
\bibliography{refs}
\bibliographystyle{numbered}

\end{document}